\documentclass[prd,twocolumn,showpacs,amsmath,amssymb]{revtex4-1}
\usepackage{siunitx}
\usepackage[sicmds,freestanding]{hepunits}
\usepackage[caption=false]{subfig}

\newcommand{\eeLScc}{\ensuremath{e^{+}e^{-}\to\Lambda\bar\Sigma^{0}+c.c.}}
\newcommand{\eeLS}{\ensuremath{e^{+}e^{-}\to\Lambda\bar\Sigma^{0}}}
\newcommand{\eeSL}{\ensuremath{e^{+}e^{-}\to\bar\Lambda\Sigma^{0}}}
\newcommand{\eeLL}{\ensuremath{e^{+}e^{-}\to\Lambda\bar\Lambda}}
\newcommand{\eeSS}{\ensuremath{e^{+}e^{-}\to\Sigma^{0}\bar\Sigma^{0}}}

\newcommand{\rmd}{{\rm d}}
\newcommand{\babar}{\textsc{BaBar}}

\newcommand{\evtgen}{\textsc{EvtGen}}
\newcommand{\conexc}{\textsc{ConExc}}
\newcommand{\photos}{\textsc{Photos}}
\newcommand{\geant}{\textsc{Geant}}
\newcommand{\fakesubsection}[1]{}
\newcommand{\fakesubsubsection}[1]{}

\usepackage{xcolor}

\usepackage{overpic}

\usepackage[utf8]{inputenc}
\usepackage[T1]{fontenc}
\usepackage{graphicx}
\usepackage{wrapfig}
\usepackage{amsmath}
\usepackage{amssymb}
\usepackage{capt-of}
\usepackage{hyperref}
\date{August 7, 2023}
\title{Measurement of the eeLScc cross sections from 2.3094 to 3.08 GeV}
\usepackage{natbib}
\begin{document}

\author{
M.~Ablikim$^{1}$, M.~N.~Achasov$^{13,b}$, P.~Adlarson$^{75}$, X.~C.~Ai$^{81}$, R.~Aliberti$^{36}$, A.~Amoroso$^{74A,74C}$, M.~R.~An$^{40}$, Q.~An$^{71,58}$, Y.~Bai$^{57}$, O.~Bakina$^{37}$
% ...
\\
\vspace{0.2cm}
(BESIII Collaboration)\\
\vspace{0.2cm} {\it
$^{1}$ Institute of High Energy Physics, Beijing 100049, People's Republic of China\\
% ...
$^{13}$ G.I. Budker Institute of Nuclear Physics SB RAS (BINP), Novosibirsk 630090, Russia\\
% ...
$^{36}$ Johannes Gutenberg University of Mainz, Johann-Joachim-Becher-Weg 45, D-55099 Mainz, Germany\\
$^{37}$ Joint Institute for Nuclear Research, 141980 Dubna, Moscow region, Russia\\
% ...
$^{40}$ Liaoning Normal University, Dalian 116029, People's Republic of China\\
% ...
$^{57}$ Southeast University, Nanjing 211100, People's Republic of China\\
$^{58}$ State Key Laboratory of Particle Detection and Electronics, Beijing 100049, Hefei 230026, People's Republic of China\\
% ...
$^{71}$ University of Science and Technology of China, Hefei 230026, People's Republic of China\\
% ...
$^{74}$ University of Turin and INFN, (A)University of Turin, I-10125, Turin, Italy; (B)University of Eastern Piedmont, I-15121, Alessandria, Italy; (C)INFN, I-10125, Turin, Italy\\
$^{75}$ Uppsala University, Box 516, SE-75120 Uppsala, Sweden\\
% ...
$^{81}$ Zhengzhou University, Zhengzhou 450001, People's Republic of China\\
\vspace{0.2cm}
% ...
$^{b}$ Also at the Novosibirsk State University, Novosibirsk, 630090, Russia\\
% ...
}
}

\begin{abstract}
The Born cross sections and effective form factors of the process \(\eeLScc\) are measured at 14 center-of-mass energy points from \SI{2.3094} to \SI{3.0800}{GeV},
based on data corresponding to an integrated luminosity of \((478.5 \pm 4.8)\ \si{pb^{-1}}\) collected with the BESIII detector.
A non-zero Born cross section is observed at the center-of-mass energy of \SI{2.3094}{GeV} with a statistical significance of more than five standard deviations,
and the cross sections at other energies are obtained with improved precision compared to earlier measurements from the \(\babar\) Collaboration.
The Born cross-section lineshape is described better by a shape with a plateau near the threshold than by a pQCD motivated functional form.
\end{abstract}

\title{Measurement of the \(\eeLScc\) cross sections at \(\sqrt{s}\) from 2.3094 to 3.0800 GeV}

\maketitle
\section{Introduction}
\label{sec:introduction}
The electromagnetic form factors~(EMFFs) are fundamental observables describing the inner, dynamical structure of hadrons and quantifying their deviation from point-like particles.
Their values can be extracted in space-like and time-like regions via scattering and annihilation processes, respectively.
In the time-like region, the EMFFs of baryons have been extensively studied in the pair-production process \(e^+e^-\to B\bar B\), where \(B\) denotes a baryon.
For spin 1/2 baryons, the Born cross section of pair-production can be parameterized in terms of the electric form factor~(FF) \(G_E\) and the magnetic FF \(G_M\) under the one-photon exchange approximation~\cite{sigma:formula},
\begin{equation}
\label{equ:sigma}
\sigma^B(s) = \frac{4\pi\alpha^2\beta\mathcal{C}}{3s}
\left[|G_M(s)|^2 + \frac {1}{2\tau}|G_E(s)|^2
\right].
\end{equation}
Here, \(\alpha\) is the fine-structure constant,
\(\beta = \sqrt{1 - 1/\tau}\) is the velocity of the baryon,
\(\tau = s/4m_B^2\),
\(s\) is the square of the center-of-mass~(c.m.) energy,
\(m_B\) is the mass of the baryon, and
\(\mathcal{C}\) is the Coulomb factor, which accounts for the \(B\bar B\) electromagnetic interactions.
The factor \(\mathcal{C}\) is unity for neutral baryons, while for charged baryon, \(\mathcal{C} = y/(1-e^{-y})\) where \(y = \pi\alpha\sqrt{1-\beta^2}/\beta\), resulting in a non-zero cross section at threshold according to Eq.~\ref{equ:sigma}.
The effective~FF \(F(s)\) is defined in terms of the moduli squared of  \(G_E\) and \(G_M\) as~\cite{formfactor:cite}
\begin{equation}F(s)
\label{equ:effectiveFF}
\equiv \sqrt{\frac{2\tau|G_M(s)|^2 + |G_E(s)|^2}{2\tau + 1}} =\sqrt{ \frac{2\tau}{2\tau+1}\frac{3s\sigma^B(s)}{4\pi\alpha^2\beta\mathcal{C}}}.
\end{equation}

In the past decades, a large number of studies have been performed to investigate the properties of baryons,
with many of them relying on the interpretation of the Born cross section of baryon pair-production.
Equation~\ref{equ:sigma} indicates that the Born cross section of the \(e^+e^-\to B\bar B\) process at the production threshold,
\(\tau = 1\),
is non-zero for charged baryon pairs due to the Coulomb correction,
and vanishes for neutral baryon pairs.
However, the measured cross sections of  \(e^+e^-\to p\bar p\)~\cite{babar:ppbar,besiii:ppbar,besiii:ppbar:2020} and \(e^+e^-\to n\bar n\)~\cite{vepp:nnbar,besiii:nnbar} processes both exhibit a flat behavior in the energy range from threshold up to about \SI{2}{GeV}.
Similar behavior near threshold is also observed in the cross sections of  \(e^+e^-\to \Lambda\bar\Lambda\)~\cite{besiii:lamlam,besiii:lamlam:ff} and \(e^+e^-\to \Lambda_c^+\bar\Lambda_c^-\)~\cite{besiii:lamclamc} production.
A similar trend is also observed for the cross sections of \(e^+e^-\to \Sigma\bar\Sigma\)~\cite{besiii:sigpsigm,besiii:sig0sig0} and \(e^+e^-\to \Xi\bar \Xi\) production~\cite{besiii:xixi,besiii:xipxim}, but larger samples are needed to reach a more reliable conclusion.
The plateau near the production threshold in the cross-section lineshape seems to be common for a variety of baryon pairs~\cite{pair:flat:behavior}.
This complicated abnormal threshold behavior has attracted great interest and stimulated many theoretical explanations, with different hypotheses being proposed such as final-state interactions~\cite{theory:final-interaction}, \(B\bar B\) bound states, vector meson resonances~\cite{theory:bbar-bound-1,theory:bbar-bound-2}, Coulomb final-state interactions or quark electromagnetic interaction and the asymmetry between attractive and repulsive Coulomb factors~\cite{theory:cf-quark-1,theory:cf-quark-2}.
The \(\babar\) experiment measured the cross section of the \(\eeLScc\) (charge conjugate) process via the initial-state radiation~(ISR) approach. The cross section
in the energy interval from threshold up to \SI{2.400}{GeV} was found 
to be \((47^{+23}_{-21}\pm 5)\)~\(\si{pb}\)~\cite{babar:lamlam:lamsig:sigsig}.
Although the uncertainty is large, the result hints at a non-zero cross section at the threshold.
This result motivates studying  \(\eeLScc\) production at and above the threshold with improved precision.

In this paper, we present the measurement of the Born cross sections and effective FFs of the \(\eeLScc\) process at c.m.~energies ranging from \SI{2.3094}{GeV}, which is 1.0 MeV above the \(\Lambda\bar\Sigma^0\) mass threshold, up to \SI{3.0800}{GeV}, with the data collected with the BESIII detector at the BEPCII collider.
A novel method is applied to measure the Born cross section near threshold
and a single-tag technique is used to improve the reconstruction efficiencies at higher energies.
In this paper, the charge conjugated process is implied.

\section{Detector and data sample}
\label{sec:detectordata}
The BESIII detector~\cite{Ablikim:2009aa} records symmetric $e^+e^-$ collisions
provided by the BEPCII storage ring~\cite{Yu:IPAC2016-TUYA01}, which operates
in the c.m. energy range from 2.00 to \SI{4.95}{GeV}.
BESIII has collected large data samples in this energy region~\cite{Ablikim:2019hff}.
The cylindrical core of the BESIII detector covers 93\% of the full solid angle and consists of a helium-based
 multilayer drift chamber~(MDC), a plastic scintillator time-of-flight
system~(TOF), and a CsI(Tl) electromagnetic calorimeter~(EMC),
which are all enclosed in a superconducting solenoidal magnet
providing a 1.0~T magnetic field~\cite{besiii:detector:desc}. The solenoid is supported by an
octagonal flux-return yoke with resistive plate counter muon-identification modules interleaved with steel.
The charged-particle momentum resolution at \SI{1}{\GeVc} is
$0.5\%$, and the $\rmd E/\rmd x$ resolution is $6\%$ for electrons
from Bhabha scattering. The EMC measures photon energies with a
resolution of $2.5\%$~($5\%$) at \SI{1}{GeV} in the barrel~(end caps)
region. The time resolution in the TOF barrel region is \SI{68}{ps}, while
that in the end-cap region is \SI{110}{ps}.

Simulated data samples produced with a \(\geant4\)-based~\cite{geant4} Monte Carlo (MC) package, which
includes the geometric and material description of the BESIII detector and the
detector response, are used to determine detection efficiencies
and to estimate backgrounds. The simulation models the beam-energy spread and initial-state radiation (ISR) in the \(e^+e^-\)
annihilations with the generator \(\conexc\)~\cite{conexc}.
All particle decays are modeled with \(\evtgen\)~\cite{evtgen-1,evtgen-2} using branching fractions taken from the Particle Data Group~(PDG)~\cite{pdg}.
Final-state radiation from charged final-state particles is incorporated with the  \(\photos\)~\cite{photos} package.

The \(\Lambda/\bar\Lambda\) and \(\Sigma^0/\bar\Sigma^0\) in the signal channel of \(\eeLS\), and dominant background channels of \(\eeLL\) and \(\eeSS\) are simulated in the \(\Lambda\to p\pi^-\)/\(\bar\Lambda\to \bar p\pi^+\) and \(\Sigma^0 \to \gamma\Lambda\)/\(\bar\Sigma^0 \to \gamma\bar\Lambda\) decay modes.

\section{Reconstruction of \(\eeLS\) at \(\sqrt{s} = 2.3094\) GeV}
\label{sec:near:threshold}
The \(\eeLS\) process is expected to dominantly produce the final state \(\gamma p \bar p\pi^+\pi^-\) at \(\sqrt{s}=\SI{2.3094}{GeV}\), which is only \(\SI{1}{MeV}\) above the kinematic threshold.
Furthermore, the decay products of \(\Lambda \to p \pi^-\) decays are close to its threshold, as are those of   \(\bar\Sigma^0\) decays.
Therefore, the particles in the final state \(\gamma p\bar p\pi^+\pi^-\) have low momenta and are unlikely to all be reconstructed by the BESIII detector.
However, low-momentum anti-protons from signal decay can interact with the material in the detector, mostly the beam pipe, and produce secondary particles. Moreover, the low momenta pions from the signal process are also mono-energetic.
Instead of demainding that all final-state particles \(\gamma p\bar p\pi^+\pi^-\) are reconstructed, an indirect search for the secondary product from the anti-proton interaction and the mono-energetic pions is employed. Similar method has been also used in Ref.~\cite{besiii:lamlam,besiii:sig0sig0}

Charged tracks detected in the MDC are required to be within $|\rm{cos\theta}|<0.93$,
where $\theta$ is defined with respect to the $z$-axis, which is the symmetry axis of the MDC.
The distance of closest approach to the interaction point~(IP) must be less than 10\,cm along the $z$-axis
and less than 1\,cm in the transverse plane.

Particle identification~(PID) for charged tracks combines measurements of the specific ionization energy loss in the MDC~(d$E$/d$x$) and the time of flight in the TOF to calculate a likelihood $\mathcal{L}(h)~(h=p,K,\pi,e)$ for each $h$ hypothesis. 
Tracks are identified as pions when the pion hypothesis has the highest likelihood value \(\mathcal{L}(\pi)>\mathcal{L}(p,K,e)\).
Events with only one \(\pi^+\) and one \(\pi^-\) reconstructed are kept for further analysis.
Considering the low momenta of the \(\Lambda\) and \(\bar\Sigma^0\), and the vertex resolution,
the \(\pi\) tracks are assumed to arise from the same vertex.
A vertex fit is applied to the two \(\pi\) tracks and the transverse distance of the vertex with respect to the beam is required to be less than \(\SI{2}{cm}\).
Both the momenta of the \(\pi^+\) and \(\pi^-\) tracks in the laboratory frame peak at around \SI{0.1}{\GeVc}.
The momentum of the \(\pi^-\) track is required to lie in the range \((0.08, 0.12)\ \si{\GeVc}\) to suppress background, and the momentum of \(\pi^+\) track is used to fit the signal contribution. 

The low-momentum anti-protons from the signal decay predominantly interact with the material in the detector, specifically the beam pipe, resulting in the production of the secondary particles.
To identify anti-protons, at least two additional charged tracks are required to originate from a common vertex.
Since the source of these secondaries are expected to be dominated by annihilations in the  beam pipe,
the transverse distance of their vertex with respect to the beam must lie between 1 and \SI{5}{cm}.

Two sources of contamination are investigated: physical  and beam-associated backgrounds.
The physical backgrounds are studied with inclusive MC samples and it is found that only the process \(\eeLL\)  survives the selection requirements.
The beam-associated background is studied with special separated-beam data taken at \(\sqrt{s} =\) 2.2324 and \SI{2.6444}{GeV},
where the beams do not collide at the IP, and is flatly distributed in $\pi^+$ momentum in the region of interest.
A data sample taken at \(\sqrt{s} =\) \SI{2.1250}{GeV},
which is below the production threshold for the signal process,
is also used to study the beam-associated background, and yields compatible results to those obtained from the   separated-beam data.

\begin{figure}
\includegraphics[width=\linewidth]{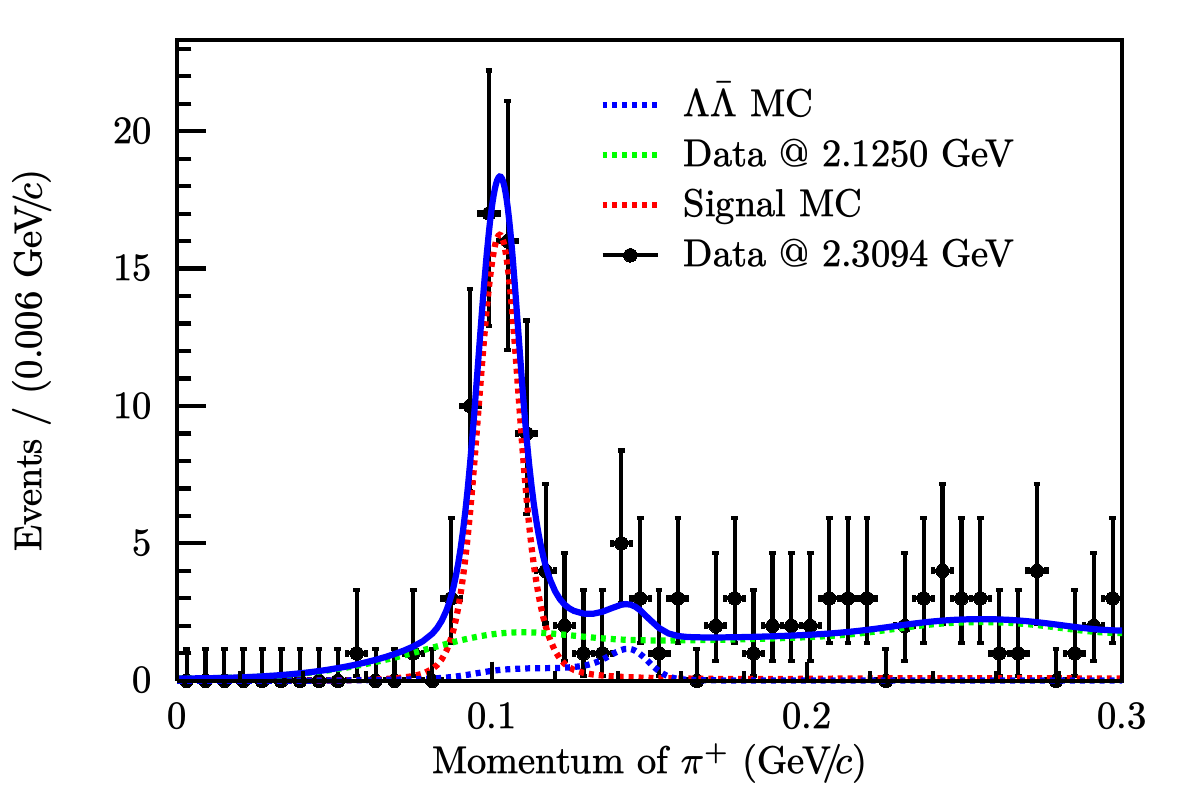}
\caption{
\label{fig:threshold:fit}
Fitted momentum distribution of reconstructed \(\pi^+\) tracks in the laboratory frame, showing the contributions of the signal together with the beam-associated background determined from data at 2.1250~GeV, and the background from \(\eeLL\) events as determined from MC simulation.
}
\end{figure}

The number of signal events is extracted by an unbinned maximum-likelihood fit to the momentum distribution of the \(\pi^+\) tracks.
A signal yield of \(49.9 \pm 8.6\) events with a statistical significance of 7.7 standard deviations is obtained by comparing the change in the log likelihood between the fits with and without the signal component, as shown in Fig.~\ref{fig:threshold:fit}.
The signal is described with an MC shape convolved with a Gaussian function describing momentum-resolution difference between data and MC simulation.
The background from the \(\eeLL\) process is described by the MC shape, and the beam-associated background is described by the shape extracted from the data taken at \(\sqrt{s} =\) \SI{2.1250}{GeV}, which has the highest integrated luminosity below the threshold of the signal process and describes the beam-associated background better.
The numbers of both background events are free in the fit.

The Born cross section of the \(\eeLS\) process is calculated by

\begin{equation}
\label{equ:sigma:cal:formula}
\sigma^B(s) = \frac{N_{\textrm{obs}}}{\mathcal{L}\cdot\epsilon\cdot (1 + \delta)\cdot \mathcal{B}},
\end{equation}
where
\(N_{\textrm{obs}}\) is the number of observed signal events in data,
\(\mathcal L\) is the integrated luminosity, \(\epsilon\) is the detection efficiency, \(1+\delta\) is the radiative correction factor due to the ISR and the vacuum polarization. Both \(\epsilon\) and  \(1+\delta\) are determined from MC simulation. 
\(\mathcal B\) is the product of the branching fractions of \(\Lambda\to p\pi^-\), \(\bar\Sigma^0\to\gamma\bar\Lambda\) and \(\bar\Lambda\to\bar p\pi^+\).

Several sources of systematic uncertainties are considered in the Born cross-section measurement at \(\sqrt{s} = \SI{2.3094}{GeV}\), which are summarized in Table~\ref{tab:threshold:sysuncern}.
The integrated luminosity is measured with 1.0\% precision~\cite{besiii:luminosity,besiii:luminosity-21250}.
The uncertainties from \(\pi^{\pm}\) tracking and PID efficiencies are determined from a control sample of \(J/\psi\to p \bar p \pi^+\pi^-\) decays, and estimated as 4.6\% and 2.0\%, respectively.
The uncertainty of \(\mathcal{B}(\Lambda \to p\pi^-)\) is 0.8\%~\cite{pdg}.
The uncertainty associated with the limited size of the signal MC sample is calculated as \(\frac {1}{\sqrt{N_{\textrm{gen}}}}\cdot \sqrt{\frac{1-\epsilon}{\epsilon}}\), where \(N_{\textrm{gen}}\) is the number of events generated in simulation.
The uncertainties introduced by selection criteria, including the requirement on the \(\pi^-\) momentum  and the transverse distance of the vertex, are studied by varying the criteria and found to be negligible.
The uncertainty from the ISR process is found to be dominated by the accuracy of \(1+\delta\) calculation in the \(\conexc\) generator and quoted as 0.5\%~\cite{conexc}. The uncertainty associated with the \(\bar p\) annihilation in the beam pipe is studied from a control sample of \(J/\psi \to p\bar p\pi^+\pi^-\) events and is determined to be 2.4\%.

The uncertainty associated with the choice of signal shape is estimated by changing the signal shape to a pure MC shape without the convolved Gaussian function.
The uncertainty associated with the beam-associated background shape is estimated by replacing the shape with  one extracted at \(\sqrt{s} =\)
2.0000 and \SI{2.1000}{GeV}.
The difference in the signal yields is taken as the uncertainty.
To study the uncertainty from the c.m.~energy spread,
 a new signal MC sample including the energy spread is generated, and the difference in \(\epsilon\cdot(1+\delta)\) is taken as the uncertainty.
All the systematic uncertainties are considered uncorrelated and summed in quadrature as listed in Table~\ref{tab:threshold:sysuncern}.

\begin{table}
\caption{
\label{tab:threshold:sysuncern}
Relative systematic uncertainties on the Born cross section measurement at \(\sqrt{s}\) = \SI{2.3094}{GeV}.}
\begin{ruledtabular}\begin{tabular}{cc}
Source & Uncertainty (\%)\\ \hline
Luminosity & 1.0 \\
\(\pi^{\pm}\) tracking & 4.6 \\
\(\pi^{\pm}\) PID & 2.0 \\
Branching fraction & 1.6\\
MC sample size &  0.8\\
\(\bar p\) annihilation & 2.4 \\
Signal shape & 2.4 \\
Background shape & 2.5 \\
Energy spread & 2.7 \\
\(1+\delta\) calculation & 0.5\\
\hline
Total & 7.4 \\
\end{tabular}\end{ruledtabular}
\end{table}

\section{Reconstruction of \(\eeLS\) at higher energies}
\label{sec:above:threshold}
At c.m.~energies from \(\sqrt {s} = 2.3864\) to \SI{3.0800}{GeV}, the final-state particles have enough momentum to be reconstructed,
but the full reconstruction still suffers from low efficiency.
Hence, a single \(\Lambda\)-tag technique is employed, i.e.,
only the \(\Lambda\) from the primary interaction is reconstructed via the decay \(\Lambda \to p\pi^-\),
and the presence of the \(\bar\Sigma^0\) is inferred from the recoil mass.

Charged tracks are reconstructed with the same method described in Sec.~\ref{sec:near:threshold},
except that
the distance of closest approach to the~IP must be less than \SI{30}{cm} along the beam direction
and less than \SI{10}{cm} in the transverse plane.
A charged track is identified as a pion (proton) when the pion (proton) hypothesis has the highest likelihood value \(\mathcal{L}(\pi)>\mathcal{L}(p,K)\) (\(\mathcal{L}(p)>\mathcal{L}(\pi,K)\)).  
Events with only one \(p\pi^-\) pair are kept for further analysis.

The \(\Lambda\) candidate is reconstructed from a \(p\pi^-\) combination satisfying a secondary-vertex fit, and having a decay length greater than twice the standard deviation of the vertex resolution. The decay length is the distance between its primary vertex and the decay point to \(p\pi^-\), where the primary vertex is the point where \(\Lambda\) is produced. The sum of \(\chi^2\) values of primary vertex fit and secondary vertex fit is required to be less than 50.
The invariant mass of the \(p\pi^-\) combination is required to be within [1.11,\SI{1.12}]{\GeVcsq}.

The \(\bar\Sigma^0\) candidate is inferred from the invariant mass of the system recoiling against the selected \(\Lambda\) candidate
\[
M_{p\pi}^{\textrm{recoil}} = \sqrt{\frac{E_{\bar\Sigma^0}^2}{c^4} - \frac{|\vec p_{e^+e^-} - \vec p_{p\pi^-}|^2}{c^2}},
\]
where \(E_{\bar\Sigma^0}\) is the expected energy of the \(\bar\Sigma^0\), i.e., \(E_{\bar\Sigma^0} = (s + m_{\bar\Sigma^0}^2c^4 - m_{\Lambda}^2c^4)/2\sqrt{s}\), and \(m_{\Lambda}\) and \(m_{\bar\Sigma^0}\) are the known masses of the labeled baryons from the PDG~\cite{pdg}. Here 
\(\vec p_{e^+e^-}\) and \(\vec p_{p\pi^-}\) are the three-momenta of the \(e^+e^-\) system and \(\Lambda\) candidate, respectively.

Potential sources of backgrounds are investigated by studying the inclusive hadronic MC samples.
The dominant background processes are \(\eeLL\) and \(\eeSS\) with final states of \(p\bar p\pi^+\pi^-\) and \(\gamma\gamma p\bar p \pi^+\pi^-\), respectively. In addition, the \(\Lambda\) produced by the decay of the \(\Sigma^0\) in the \(\eeSL\) process also contributes to the background.
Besides background from \(\Lambda\) recoil, there is also combinatorial background which can be estimated using the \(\Lambda\) sidebands.
The \(\Lambda\) sidebands defined as \(M_{p\pi}\) $\in$ (1.095, 1.105) \(\si{\GeVcsq}\) and \(M_{p\pi}\) $\in$ (1.125, 1.135) \(\si{\GeVcsq}\) are inspected, and indicate that there is a non-zero level of combinatorial background, but no significant peaking  contamination.

\begin{figure}
\includegraphics[width=\linewidth]{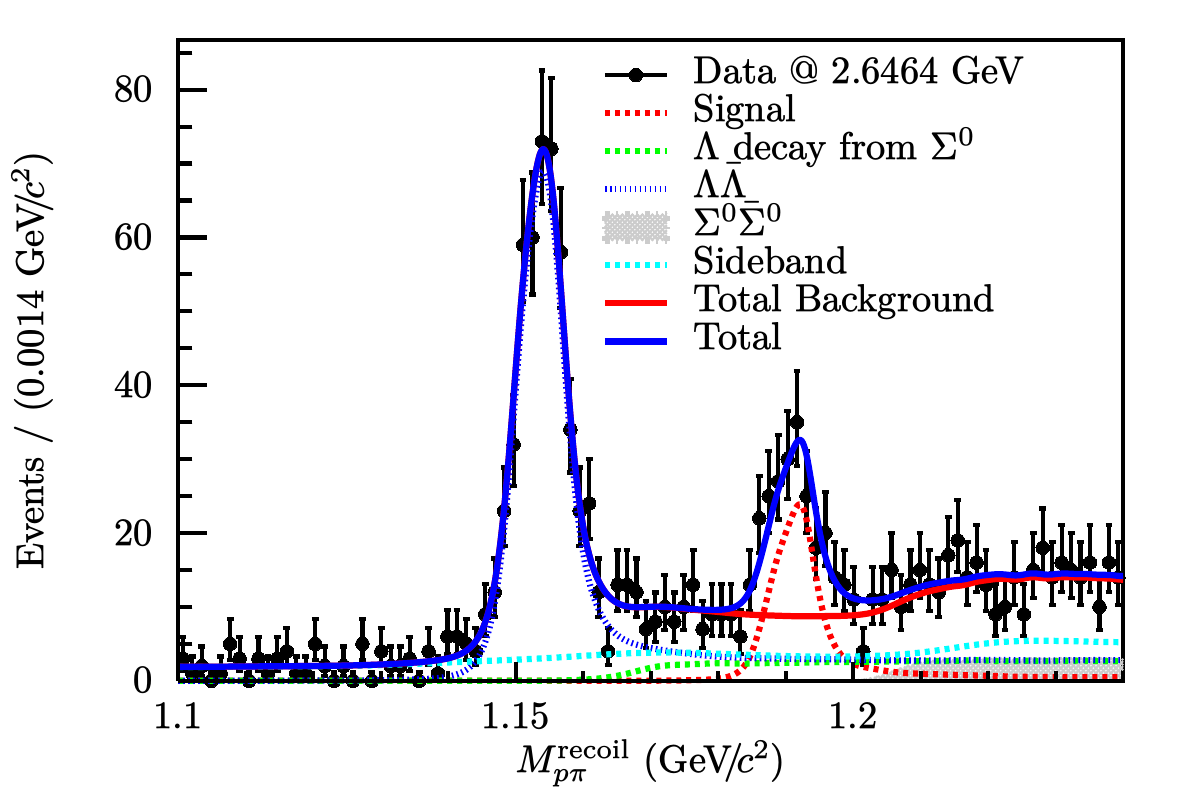}
\caption{
\label{fig:nonthreshold:fit}
Fitted distribution of \(M_{p\pi}^{\textrm{recoil}}\) at \(\sqrt{s} = \SI{2.6464}{GeV}\).
The black dots with error bars are the data,
The blue solid line is the fit result, and the red solid line is the overall background.
The red dashed line represents the signal,
the blue dashed line represents the \(\Lambda\bar\Lambda\) background,
the green dashed line represents the background where \(\Lambda\) decays from \(\Sigma^0\),
the cyan dashed line represents the background extracted from \(M_{p\pi}\) sideband,
and the gray filled curve represents the \(\Sigma^0\bar\Sigma^0\) background.
}
\end{figure}

The  \(\eeLS\) signal yields at each energy point are extracted by an unbinned maximum-likelihood fit to the \(M_{p\pi}^{\textrm{recoil}}\) spectrum, an example of which is shown at \(\sqrt{s} = \SI{2.6444}{GeV}\)  in Fig.~\ref{fig:nonthreshold:fit}.
The signal is described with the shape obtained from MC, convolved with a Gaussian function to compensate for possible mass-resolution difference between data and MC simulation.
The background is modeled with the MC shapes of the \(\eeLL\) process, the \(\eeSS\) process, the \(\Lambda\) from \(\Sigma^0\) decay in the signal process and the sideband shape from experimental data.
The MC shape of the \(\eeLL\) process is convolved with the same Gaussian function as the signal shape.
The MC shape of \(\Lambda\) from \(\Sigma^0\) in the \(\eeSL\) process is fixed according to the signal MC sample.
The Born cross section is determined using Eq.~\ref{equ:sigma:cal:formula},
where \(\mathcal{B}\) is the branching fraction of the decay \(\Lambda \to p\pi^-\).

Various sources of systematic uncertainties have been considered in Born cross-section measurements in this energy regime, with a summary presented in Table~\ref{tab:nonthreshold:sysuncern}. The uncertainties associated with the  luminosity measurement, branching fraction knowledge, MC sample size, signal lineshape and \(1+\delta\) calculation are assigned following the same procedure as described in the previous section.
The uncertainty associated with the reconstruction of the \(\Lambda\) is determined from the control samples \(J/\psi\to \bar pK^+\Lambda\) and \(J/\psi \to \Lambda\bar\Lambda\), with a similar method as used in Ref.~\cite{lam:uncern}.
To estimate the uncertainty coming from the knowledge of  the angular distribution of the \(\Lambda\) baryon, the analysis is repeated with the two extremes of the angular distributions \((1 + \cos^2 \theta)\) and \((1-\cos^2\theta)\)~\cite{data:ana:book}. The difference in the resulting efficiencies divided by \(\sqrt{12}\) is taken as the uncertainty.
Alternative fits are performed to study the uncertainties associated with the fit shapes. These include changing the default signal shape to a pure MC shape without the convolved Gaussian function, varying the regions of sideband background to only \([1.095, 1.105]\ \si{\GeVcsq}\) or \([1.125, 1.135]\ \si{\GeVcsq}\), replacing the input lineshape of \(\Lambda\bar\Lambda\) background MC sample with one extracted from the cross section of the \(\Lambda\) pair production at BESIII experiment~\cite{besiii:lamlam,besiii:lamlam:ff} instead of the fitted result from BaBar experiment, and fixing the background shape of the \(\Sigma^0\bar\Sigma^0\) process according to the integrated luminosity and its cross section. 
Any potential bias arising from the requirement on the \(\chi^2\) of the vertex fit is investigated by varying the boundaries from 20 to 100,
and that from the  mass-window requirement of the \(\Lambda\) candidate is estimated by varying the left boundary from 1.109 to \(\SI{1.112}{\GeVcsq}\) and the right boundary from 1.118 to \(\SI{1.122}{\GeVcsq}\). Both of these contributions are found to be negligible, and no uncertainties are assigned. 
All the systematic uncertainties are considered uncorrelated and summed in quadrature.

\onecolumngrid

\begin{table*}[htb!]
  \caption{
  \label{tab:nonthreshold:sysuncern}
  Relative systematic uncertainties (in \%) in the cross section measurement
 for each c.m. energy (\(\sqrt{s}\)) above \SI{2.3864}{GeV}:
 the uncertainty associated with luminosity ($\mathcal L$), \(\Lambda\) reconstruction~(\(\Lambda\)), MC sample size~(\(\epsilon\)), branching fraction~(\(\mathcal{B}\)), \(\Lambda\) angular distribution~(Angle), signal shape~(Sig), sideband shape~(Sideband), \(\Lambda\bar\Lambda\) background shape~(\(\Lambda\bar\Lambda\) shape), \(\Sigma^{0}\bar\Sigma^{0}\) background shape~(\(\Sigma^0\bar\Sigma^0\) shape), signal input lineshape and \(1+\delta\) calculation~(\(1+\delta\)).
    The last column gives the total systematic uncertainty.
  }
\begin{ruledtabular}\begin{tabular}{ c ccccc ccc ccc}
$\sqrt{s}(\si{GeV})$ &
$\mathcal{L} $ & $\Lambda$ & $\epsilon$ & $\mathcal{B}$ &  Angle & $\textrm{Sig}$ & $\textrm{Sideband}$ &  $\Lambda\bar\Lambda$ shape &
$\Sigma\bar\Sigma^{0}$ shape &   $1+\delta$ &
Total \\
\hline

2.3864 &
1.0 & 2.6 & 0.9 & 0.8 & 7.8 &
1.7 & 7.7 & 0.9 &
0.0 & 0.5 &
11.5\\

2.3960 &
1.0 & 2.6 & 0.8 & 0.8 & 7.4 &
1.3 & 1.1 & 0.5 &
0.0 & 0.5 &
8.2\\

2.5000 &
1.0 & 2.6 & 0.5 & 0.8 & 6.7 &
2.5 & 2.5 & 3.7 &
2.3 & 0.5 &
9.3\\

2.6444 &
1.0 & 3.3 & 0.4 & 0.8 & 5.7 &
0.7 & 3.5 & 7.9 &
0.1 & 0.5 &
10.9\\

2.6464 &
1.0 & 3.3 & 0.4 & 0.8 & 6.1 &
0.3 & 6.8 & 8.3 &
0.6 & 0.5 &
12.9\\

  2.7000 &
1.0 & 3.3 & 0.4 & 0.8 & 5.7 &
0.3 & 6.8 & 8.3 &
0.6 & 0.5 &
12.7\\

2.8000 &
1.0 & 3.3 & 0.4 & 0.8 & 5.6 &
2.6 & 5.5 & 7.3 &
4.6 & 0.5 &
12.5\\

2.9000 &
1.0 & 3.3 & 0.5 & 0.8 & 5.4 &
0.6 & 2.9 & 0.9 &
1.5 & 0.5 &
7.3\\

2.9500 &
1.0 & 3.1 & 0.5 & 0.8 & 5.8 &
0.4 & 13.8 & 0.5 &
0.3 & 0.5 &
15.3\\

2.9810 &
1.0 & 3.1 & 0.5 & 0.8 & 5.7 &
0.2 & 0.1 & 2.2 &
10.2 & 0.5 &
12.4\\

3.0000 &
1.0 & 3.1 & 0.5 & 0.8 & 5.8 &
0.1 & 3.3 & 1.6 &
1.4 & 0.5 &
7.8\\

3.0200 &
1.0 & 3.1 & 0.5 & 0.8 & 5.9 &
3.2 & 3.9 & 1.7 &
2.9 & 0.5 &
9.1\\

3.0800 &
1.0 & 3.1 & 0.5 & 0.8 & 5.7 &
0.5 & 4.0 & 2.6 &
4.4 & 0.5 &
9.3\\

\end{tabular}
\end{ruledtabular}
\end{table*}

\twocolumngrid

\begin{table*}[htb!]
\caption{
\label{tab:sigma:effectiveFF}
The luminosity ($\mathcal L$), signal yield ($N_{\rm obs}$), detection efficiency ($\epsilon$), radiative  correction factor ($1+\delta$), obtained Born cross section ($\sigma^{\rm B}$) and effective FF ($F(s)$) for the $e^+e^-\to \Lambda \bar \Sigma^0$ process at
each c.m. energy (\(\sqrt{s}\)).
The first uncertainties are statistical and the second ones are systematic. The uncertainties for the signal yields are statistical only.}
\begin{ruledtabular}\begin{tabular}{ccc cccc}
\(\sqrt{s}\) (GeV) & \(\mathcal{L}\) \((\si{pb^{-1}})\) & \(N_{\textrm{obs}}\) & \(\epsilon\) & \(1+\delta\) & \(\sigma^{\textrm{B}}\) (pb) & \(F(s)\)\\
\hline
 2.3094 & 21.1 & $\phantom{0}49.9 \pm \phantom{0}8.6$ & 0.127 & 0.627 &
    $72.9 \pm 12.6 \pm 5.4$ &
    $0.315 \pm 0.027 \pm 0.012$\\

  2.3864 & 22.5 & $124.2 \pm 11.8$ & 0.107 & 0.886 &
    $91.4 \pm \phantom{0}8.7 \pm 10.6$ &
    $0.128 \pm 0.006 \pm 0.007$\\

  2.3960 & 66.9 & $447.2 \pm 22.1$ & 0.134 & 0.896 &
    $87.3 \pm \phantom{0}4.3 \pm 7.1$ &
    $0.122 \pm 0.003 \pm 0.005$\\

  2.5000 & 1.10 & $\phantom{0}\phantom{0}7.6 \pm \phantom{0}2.8$ & 0.301 & 0.985 &
    $36.6 \pm 13.3 \pm 3.4$ &
    $0.070 \pm 0.013 \pm 0.003$\\

  2.6444 & 33.7 & $183.8 \pm 16.0$ & 0.338 & 1.095 &
    $23.1 \pm \phantom{0}2.0 \pm 2.5$ &
    $0.053 \pm 0.002 \pm 0.003$\\

  2.6464 & 34.0 & $163.2 \pm 15.4$ & 0.338 & 1.096 &
    $20.3 \pm \phantom{0}1.9 \pm 2.6$ &
    $0.050 \pm 0.002 \pm 0.003$\\

  2.7000 & 1.03 & $\phantom{0}\phantom{0}0.0 \pm \phantom{0}2.3$ & 0.344 & 1.140 &
    $\phantom{0}0.0 \pm \phantom{0}8.9 \pm 0.0$ &
    $0.000 \pm 0.033 \pm 0.000$\\

  2.8000 & 1.01 & $\phantom{0}\phantom{0}1.8 \pm \phantom{0}1.6$ & 0.333 & 1.231 &
    $\phantom{0}6.8 \pm \phantom{0}6.0 \pm 0.9$ &
    $0.029 \pm 0.013 \pm 0.002$\\

  2.9000 & 106 & $272.0 \pm 18.8$ & 0.312 & 1.337 &
    $\phantom{0}9.6 \pm \phantom{0}0.7 \pm 0.7$ &
    $0.034 \pm 0.001 \pm 0.001$\\

  2.9500 & 15.9 & $\phantom{0}25.9 \pm \phantom{0}7.1$ & 0.295 & 1.397 &
    $\phantom{0}6.2 \pm \phantom{0}1.7 \pm 0.9$ &
    $0.028 \pm 0.004 \pm 0.002$\\

  2.9810 & 16.1 & $\phantom{0}30.9 \pm \phantom{0}6.1$ & 0.297 & 1.436 &
    $\phantom{0}7.0 \pm \phantom{0}1.4 \pm 0.9$ &
    $0.030 \pm 0.003 \pm 0.002$\\

  3.0000 & 15.9 & $\phantom{0}30.6 \pm \phantom{0}6.3$ & 0.289 & 1.461 &
    $\phantom{0}7.1 \pm \phantom{0}1.5 \pm 0.6$ &
    $0.030 \pm 0.003 \pm 0.001$\\

  3.0200 & 17.3 & $\phantom{0}27.3 \pm \phantom{0}5.9$ & 0.287 & 1.486 &
    $\phantom{0}5.8 \pm \phantom{0}1.2 \pm 0.5$ &
    $0.027 \pm 0.003 \pm 0.001$\\

  3.0800 & 126 & $136.9 \pm 14.4$ & 0.254 & 1.511 &
    $\phantom{0}4.4 \pm \phantom{0}0.5 \pm 0.4$ &
    $0.024 \pm 0.001 \pm 0.001$\\

  \end{tabular}\end{ruledtabular}
\end{table*}

\begin{figure*}[htb!]
\begin{Overpic}[grid=false]{%
\includegraphics[width=.45\linewidth]{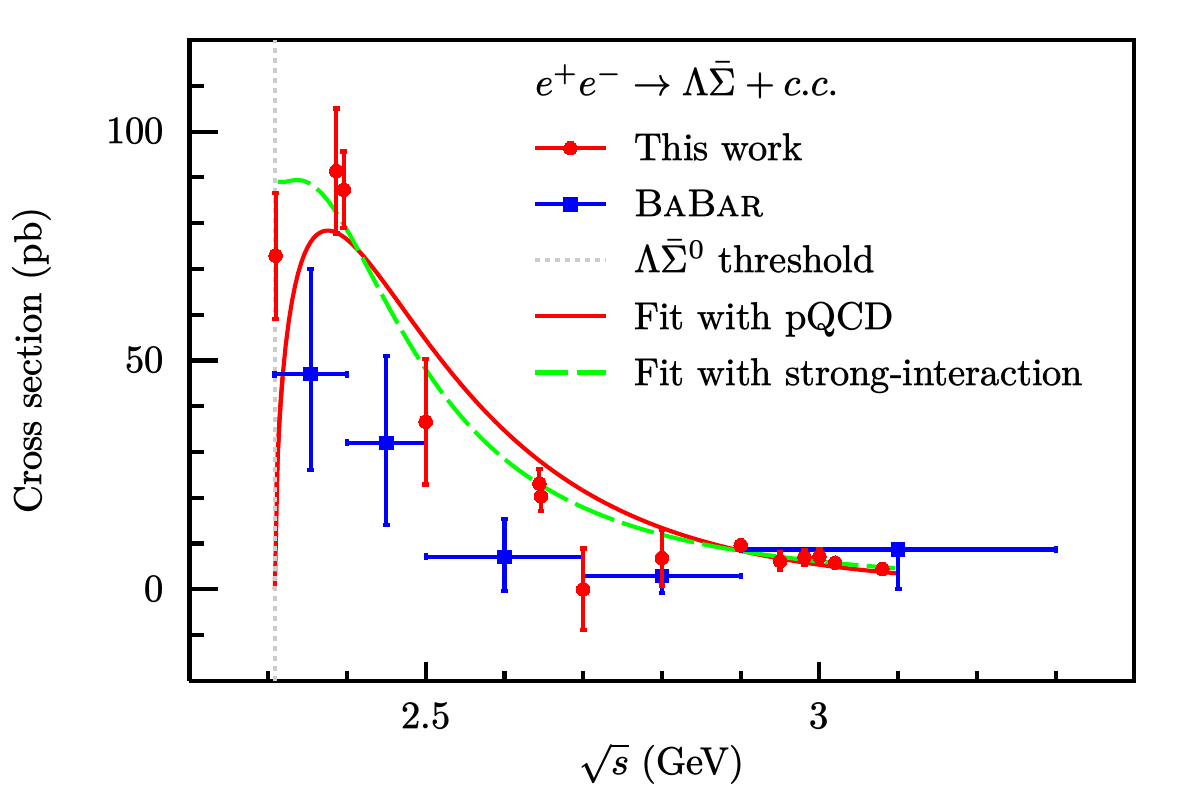}}%
\put(28,56){(a)}
\end{Overpic}%
\begin{Overpic}[grid=false]{    %
\includegraphics[width=.45\linewidth]{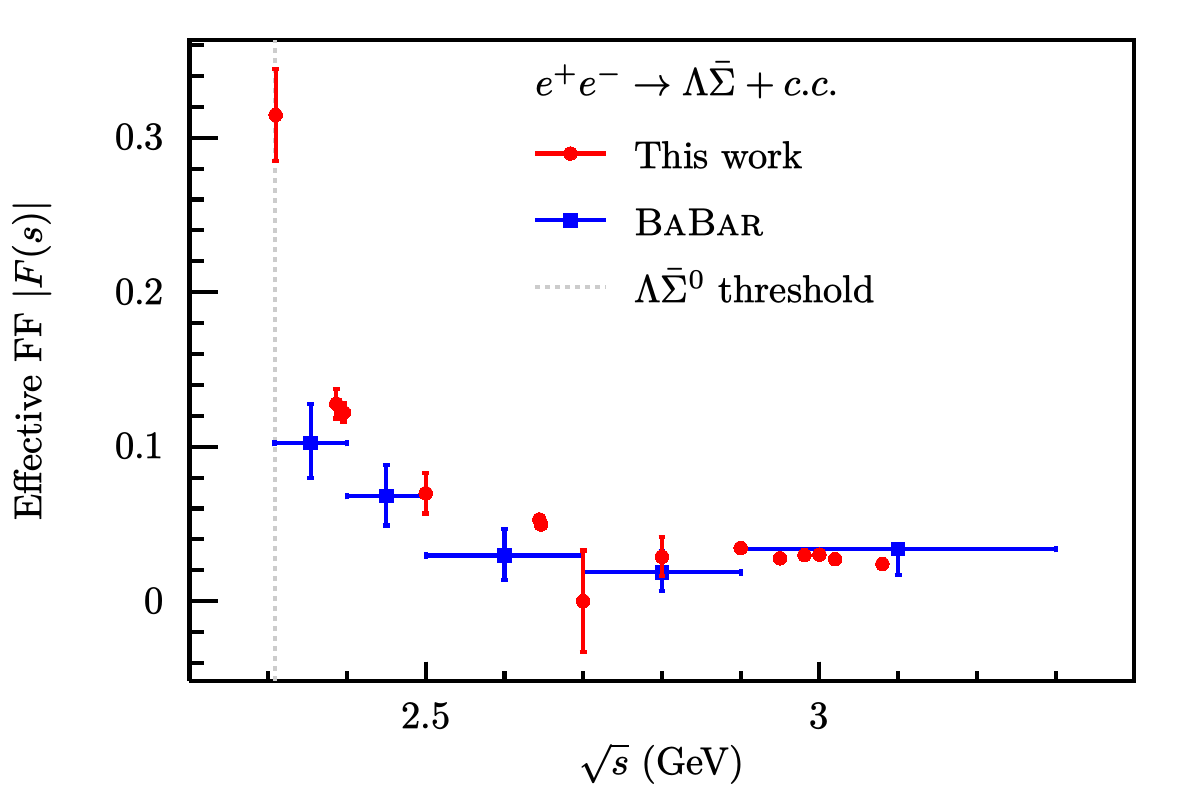}}%
\put(28,56){(b)}
\end{Overpic}%
\caption{
\label{fig:lineshape:effectiveFF}
(a)~The Born cross sections and (b)~effective FFs for the \(\eeLScc\) process from this analysis shown with fits to the pQCD model ( Eq.~\ref{equ:pqcd:formula}) and a model with strong-interaction effects included (Eq.~\ref{equ:ppbar:fit}).  Also shown are the results obtained by  \(\babar\).
}
\end{figure*}

\section{Results and conclusion}
\label{sec:results}
For the \(\eeLS\) process, the expressions of Eq.~\ref{equ:sigma} and Eq.~\ref{equ:effectiveFF}
need to be modified by the substitutions,
\begin{gather}
\label{equ:lamsig:beta}
\beta = \sqrt{1 - \frac{2(m_{\bar\Sigma^0}^2 + m_{\Lambda}^2)}{s} + \frac{(m_{\bar\Sigma^0}^2 - m_{\Lambda}^2)^2}{s^2}},\\
\tau = \frac{s}{(m_{\Lambda} + m_{\bar\Sigma^0})^2}.
\end{gather}
The resulting Born cross sections and the effective FFs are summarized in Table~\ref{tab:sigma:effectiveFF},
and a comparison between the results of  this work and those of \(\babar\) is illustrated in Fig.~\ref{fig:lineshape:effectiveFF}.

A perturbative QCD~(pQCD) motivated energy power function~\cite{sigma:pqcd:formula}, given by
\begin{equation}
\label{equ:pqcd:formula}
\sigma^{\textrm{B}}(s) = \frac{c_0\cdot \beta \cdot \mathcal{C}}{(\sqrt{s} - c_1)^{10}},
\end{equation}
is used to fit the lineshape, where \(c_0\) and \(c_1\) are free parameters and
the Coulomb correction factor \(\mathcal{C}\) is 1 for a neutral channel.
Figure~\ref{fig:lineshape:effectiveFF}(a) shows the fit result, 
with \(c_0 = (9.94 \pm 3.91) \times 10^4\ \si{pb\cdot GeV^{10}}\), \(c_1 = (0.97 \pm 0.07)\ \si{GeV}\) and fit quality \(\chi^2/ndof = 41.0 / 12 = 3.42\), where \(ndof\) is the number of degrees of freedom.
From the fit quality, the non-zero cross section at threshold does not fit to the pQCD model.

Inspired by the nucleon pair-production cross section and its plateau near threshold region~\cite{besiii:ppbar:2020}, another prediction of the Born cross section is used to describe the lineshape, which takes into account strong-interaction effects near the threshold,

\begin{equation}
\label{equ:ppbar:fit}
\sigma^{\textrm{B}}(s) =
\frac{e^{a_0}\pi^2\alpha^3}
		{s\left[1 - e^{-\frac{\pi\alpha_s(s)}{\beta(s)}}\right]\left[ 1 + \left( \frac{\sqrt s - (m_\Lambda + m_{\bar\Sigma^0}) }{a_1}\right)^{a_2} \right]}.
\end{equation}
Here \(a_0\), \(a_1\), \(a_2\) are free parameters, \(a_0\) is the normalization constant,
\(a_1\) is the QCD parameter near threshold, \(a_2\) is a power-law parameter related to the number of valence quarks, \(\alpha\) is the electromagnetic coupling constant and
\(\alpha_s(s)\) is the strong running coupling constant,
\begin{equation}
\label{equ:ppbar:alphas}
	\alpha_s(s) = \left[\frac {1}{\alpha_s(m_{\mathcal Z}^2)} + \frac {25}{12\pi}\ln \left(\frac {s}{m_{\mathcal Z}^2} \right)\right]^{-1},
\end{equation}
where \(m_{\mathcal Z} = \SI{91.1876}{\GeVcsq}\) is the \(Z\) boson mass and \(\alpha_s(m_{\mathcal Z}^2) = 0.11856\).
Figure~\ref{fig:lineshape:effectiveFF}(a) also shows the fit result, with
\(a_0 = 19.79 \pm 0.10\),
\(a_1 = (0.17 \pm 0.02)\ \si{GeV}\),
\(a_2 = 1.99 \pm 0.16\) and
\(\chi^2/ndof = 11.8 / 11 = 1.07\).
The description of Eq.~\ref{equ:ppbar:fit} that includes strong-interaction effects  gives a better fit quality than the pQCD prediction and the inflection point of the plateau near threshold is roughly \SI{100}{MeV} higher than threshold. Figure~\ref{fig:lineshape:effectiveFF}(b) shows the effective FFs obtained in this work and previous measurement at \babar{}. Except for the c.m. emergy of \SI{2.3094}{GeV}, our measured results are consistent with earlier results from \babar{}, with improved precision.

\section{Summary}
\label{sec:summary}
Based on a total integrated luminosity of \(\SI{478.5}{pb^{-1}}\) \(e^+e^-\) collision data collected with the BESIII detector,
the Born cross sections and effective form factors of the \(\eeLS\) process have been determined at c.m.~energies ranging from 2.3094 up to \SI{3.0800}{GeV}.
At \(\sqrt s = \SI{2.3094}{GeV}\), which is approximately 1 MeV above the threshold,
the signal process is identified by the primary pion from the signal decay and secondary tracks from the interaction of anti-proton with beam pipe.
A non-zero Born cross section is found
with a statistical significance greater than 5 standard deviations
and measured to be \((72.9 \pm 12.6 \pm 5.4)\) \(\si{pb}\),
where the first uncertainty is statistical and the second is systematic.
At other energies, a single-tag technique is employed by tagging the primary \(\Lambda\) alone to optimize the detection efficiency.
The Born cross sections at these energies are in good agreement with those of \babar{}, but with improved precision.
Fits with pQCD assumption and the plateau near threshold are performed on the lineshape of the Born cross sections, and it is found that the latter gives a better description of the data. The measured effective FFs are consistent with \babar{}'s results for the c.m. energies above \SI{2.3094}{GeV}.

\section{Acknowledgments}
\label{sec:ack}
The BESIII Collaboration thanks the staff of BEPCII and the IHEP computing center and the supercomputing center of USTC for their strong support. This work is supported in part by National Key R\&D Program of China under Contracts Nos. 2020YFA0406300, 2020YFA0406400; National Natural Science Foundation of China (NSFC) under Contracts Nos. 12105276, 11335008, 11625523, 12035013, 11705192, 11950410506, 12061131003, 12122509, 11635010, 11735014, 11835012, 11935015, 11935016, 11935018, 11961141012, 12022510, 12025502, 12035009, 12192260, 12192261, 12192262, 12192263, 12192264, 12192265, 12221005, 12235017; Joint Large-Scale Scientific Facility Funds of the NSFC and CAS under Contracts Nos. U1732263, U1832103, U2032111; the Chinese Academy of Sciences (CAS) Large-Scale Scientific Facility Program; the CAS Center for Excellence in Particle Physics (CCEPP); CAS Key Research Program of Frontier Sciences under Contracts Nos. QYZDJ-SSW-SLH003, QYZDJ-SSW-SLH040; 100 Talents Program of CAS; The Institute of Nuclear and Particle Physics (INPAC) and Shanghai Key Laboratory for Particle Physics and Cosmology; ERC under Contract No. 758462; European Union's Horizon 2020 research and innovation programme under Marie Sklodowska-Curie grant agreement under Contract No. 894790; German Research Foundation DFG under Contracts Nos. 443159800, 455635585, Collaborative Research Center CRC 1044, FOR5327, GRK 2149; Istituto Nazionale di Fisica Nucleare, Italy; Ministry of Development of Turkey under Contract No. DPT2006K-120470; National Research Foundation of Korea under Contract No. NRF-2022R1A2C1092335; National Science and Technology fund of Mongolia; National Science Research and Innovation Fund (NSRF) via the Program Management Unit for Human Resources \& Institutional Development, Research and Innovation of Thailand under Contract No. B16F640076; Polish National Science Centre under Contract No. 2019/35/O/ST2/02907; The Swedish Research Council; U. S. Department of Energy under Contract No. DE-FG02-05ER41374

\bibliography{main}{}
\end{document}